\documentclass[conference,10pt]{IEEEtran}

\usepackage{psfrag}
\usepackage{amsmath}
\usepackage{amssymb}
\usepackage{amsfonts}
\usepackage{graphicx}
\usepackage{float}
\usepackage{graphics}
\usepackage{graphicx}
\usepackage{theorem}
\usepackage{euscript}

\begin{document}

\title{Distributed MIMO Systems with Oblivious Antennas}
\author{\authorblockN{Osvaldo Simeone} 
\authorblockA{CWCSPR, ECE Department\\   New Jersey Institute of Technology\\
Newark, NJ 07102\\  
osvaldo.simeone@njit.edu} \and 
\authorblockN{Oren Somekh and H. Vincent
Poor} 
\authorblockA{Department of Electrical Engineering\\
Princeton University\\
Princeton, NJ 08544\\
$\{$orens, poor$\}$@princeton.edu}\and 
\authorblockN{Shlomo Shamai (Shitz)} 
\authorblockA{Department of Electrical Engineering\\
Technion\\
Haifa, 32000, Israel\\
sshlomo@ee.technion.ac.il}}
\maketitle

\begin{abstract}
A scenario in which a single source communicates with a single destination
via a distributed MIMO transceiver is considered. The source operates each
of the transmit antennas via finite-capacity links, and likewise the
destination is connected to the receiving antennas through
capacity-constrained channels. Targeting a nomadic communication scenario,
in which the distributed MIMO\ transceiver is designed to serve different
standards or services, transmitters and receivers are assumed to be
oblivious to the encoding functions shared by source and destination.
Adopting a Gaussian symmetric interference network as the channel model (as
for regularly placed transmitters and receivers), achievable rates are
investigated and compared with an upper bound$.$ It is concluded that in
certain asymptotic and non-asymptotic regimes obliviousness of transmitters
and receivers does not cause any loss of optimality.
\end{abstract}



%


\section{Introduction\label{sec_intro}}

MIMO systems implemented via \textit{distributed antennas}, connected via a
wireless or wired backbone, have been recently advocated as a viable
solution to provide multiplexing, array and micro- or macro-diversity gains
in infrastructure or mesh/ ad hoc networks (see, e.g., \cite{goldsmith}-\cite%
{katz} and references therein). In such systems with non-colocated antennas,
the main challenge in realizing the full gains of MIMO systems is the
efficient use of the channel resources needed to coordinate the
participating antennas at the transmit and/ or receive sides into effective
multi-antenna arrays. These channel resources can be either \textit{in-band,}
that is, in the same time and frequency band of the main end-to-end
transmission \cite{goldsmith} \cite{katz}, or \textit{out-of-band}, i.e.,
over orthogonal channels, possibly realized via wired connections or
different wireless radio interfaces \cite{somekh-review} \cite{shamai-review}%
. Moreover, the feasibility of different transmission schemes depends on the
amount of information that the available transceivers have regarding the
encoding functions shared by the sources and destinations of the transmitted
data. For example, decode-and-forward-type schemes require full knowledge of
the codebooks used to encode the received data, while compress-and-forward
or amplify-and-forward strategies do not have this requirement.


\begin{figure}
\centering
 \includegraphics[scale=0.55]{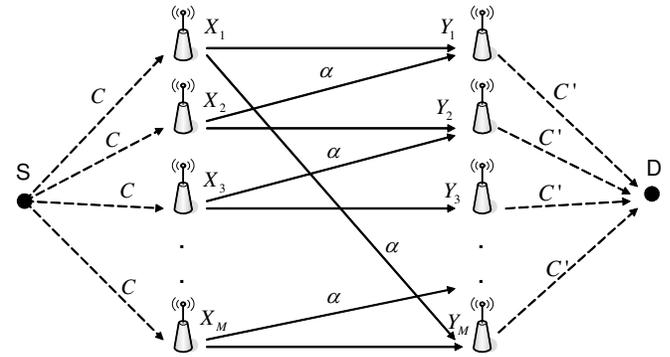}
 \caption{System model: a nomadic single source \textrm{S} communicates to a single destination \textrm{D }over a symmetric interference channel where transmitters and receivers are oblivious to the encoding functions employed by\textrm{\ S} and \textrm{D}.}
 \label{fig1}
\end{figure}

Basic $2\times 2$ distributed MIMO systems with full codebook information
and with in-band signalling between transmit antennas, at one end, and
receive antennas, at the other, were considered in \cite{goldsmith} and \cite%
{host} for half and full-duplex transceivers, respectively. Reference \cite%
{sanderovich-new} studies a general MIMO system with infinite-capacity
out-of-band links connecting the transmit-side antennas and finite-capacity
links at the receive side, where antennas are assumed to be either
codebook-unaware (i.e., \textit{oblivious}) or informed. Another line of
work that is of interest here deals with distributed-antenna transmitters or
receivers in a cellular scenario in the presence of non-ideal out-of-band
links connecting the cooperating (transmit or receive) antennas (base
stations), which can be either oblivious or not \cite{shamai-review}.

In this paper, we consider the scenario depicted in Fig. \ref{fig1} in which
a single source $\mathrm{S}$ has data to communicate to a remote destination 
$\mathrm{D.}$ Communication takes place via a distributed $M\times M$\ MIMO
system constructed by connecting the source to the $M$ transmitting antennas
through (equal) finite-capacity links and likewise the $M$ receiving
antennas to the destination. The finite-capacity links are assumed to be
orthogonal among themselves and out-of-band. This is the case when the
source and destination are connected to the transmitters and receivers,
respectively, via a wired backbone or via orthogonal wireless interfaces.
Targeting a scenario where the infrastructure of transmitting and receiving
antennas is meant to serve different communication standards, we assume that
transmitters and receivers are \textit{oblivious} to the encoding function
shared by source \textrm{S} and destination \textrm{D }as in\textrm{\ }\cite%
{shamai-review} \cite{sanderovich-new}. Our interest is in obtaining
analytical insights into the role of finite capacities $C$ and $C^{\prime }$
at the transmit and receive sides, respectively, on the performance of the
distributed MIMO\ system (see Fig. \ref{fig1}). To this end, we adopt a
simplified channel model for the MIMO channel between transmitters and
receivers that corresponds to a Gaussian interference network described by a
single parameter $\alpha $, as shown in Fig. \ref{fig1}. Beside allowing
analytical tractability, this channel is a variant of the Wyner model for
infrastructure networks \cite{wyner} that has been studied in a number of
works (see \cite{somekh-review} and \cite{shamai-review} for a review). An
upper bound and achievable rates are derived. It is shown that, in certain
asymptotic and non-asymptotic regimes, no loss of optimality is incurred in
designing the system for nomadic applications (i.e., assuming oblivious
transmitters and receivers).

\section{System model \label{sec sys}}

A source \textrm{S} is connected via finite-capacity (error-free) links of
capacity $C$ [bit/ symbol] to $M$ distributed transmitters. Each transmitter
has power constraint $P$. The source aims at communicating with a remote
destination \textrm{D, }which is in turn connected to $M$ distributed
receivers via links of capacity $C^{\prime }.$ Targeting nomadic
applications, transmitters and receivers are assumed to be \textit{oblivious}
to the encoding function shared by source \textrm{S} and destination \textrm{%
D. }More specifically: (\textit{i}) Each transmitter is equipped with an
independently generated standard complex Gaussian codebook of size $2^{nC}$ (%
$n$ is the length of the transmission block) with average power $P$, which
is known to the source \textrm{S}. These $M$ codebooks can be obtained by 
\textrm{S} via, e.g., a local public database. Through the finite-capacity
link, the source selects which codeword in the codebook should be sent by
each antenna in a given transmission block. In other words, no processing is
carried out at the transmitters, except simple \textit{mapping between the
index received by the source and the codebook}; (\textit{ii}) Each receiver
is unaware of the processing carried out at the source and of the codebooks
of the transmitters, and merely performs quantization of the received
signals, which are then relayed to the destination; (\textit{iii}) The
destination \textrm{D} is assumed to be aware of the quantization scheme
used at the receivers. Moreover, we consider both cases in which the
destination knows and does not know the codebooks of the transmitters.
Finally, perfect block and symbol synchronization is assumed.

The complex Gaussian channel between transmitters and receivers is described
by a Gaussian interference network as in Fig. \ref{fig1}, which is further
described by an interference parameter $\alpha \in \lbrack 0,1]$. This
channel corresponds to the circulant version of the Wyner model \cite{wyner}
for cellular networks considered in various works (see \cite{somekh-review}\ 
\cite{shamai-review} for reviews). Accordingly, the received signal at any
given time instant by the $m$th receiver is given by 
\begin{equation}
Y_{m}=X_{m}+\alpha X_{[m-1]}+Z_{m},  \label{rx_signal}
\end{equation}%
where $X_{m}$ is the complex symbol transmitted by the $m$th transmitter, $%
[m-1]$ represents the modulo-$M$ operation, and $Z_{m}$ is complex Gaussian
noise with unit power ($Z_{m}\sim \mathcal{CN}(0,1)$). The per-transmitter
input power constraint requires $E[|X_{m}|^{2}]=P.$ Parameters $\alpha ,$ $C$
and $C^{\prime }$ are assumed to be known by all the involved nodes. In
order to obtain compact results, we will focus on the case $M\rightarrow
\infty .$ Results with finite $M$ can be easily inferred by using the
circular structure of the channel model at hand and the corresponding
circularity of the channel matrix, based on the standard arguments (see \cite%
{somekh-review} for a review and a discussion on the validity of this
asymptotic analysis)$.$ Finally, we normalize the achievable rate from 
\textrm{S} and \textrm{D }to the number of transmit and receive antennas $M$
and definite it as $R$ [bit/ (symbol $\times $ antenna)]. Results will be
stated here without formal proof. The reader is referred to \cite{fullpaper}
for proofs and further discussion.

\section{Preliminaries}

In this section, we review some basic definitions and establish a reference
result. At first, we recall that for $M\rightarrow \infty $ and perfectly
cooperating receivers ($C^{\prime }\rightarrow \infty $), the signal (\ref%
{rx_signal}) received by the destination of the network in Fig. \ref{fig1}
can be interpreted in the spatial domain as an inter-symbol-interference
channel with impulse response $h_{m}=\delta _{m}+\alpha \delta _{m-1}$ ($%
\delta _{m}$ is the Kronecker delta function) and frequency response \cite%
{wyner} (see also \cite{somekh-review}):%
\begin{equation}
|H(f)|^{2}=1+\alpha ^{2}+2\alpha \cos (2\pi f).  \label{channel}
\end{equation}%
We then present two basic definitions and related results.

\textit{Definition 1}: We define the waterfilling power spectral density
with respect to the sum-power constraint $P$ and the SNR power spectral
density $\rho (f)$ as $S_{WF}(f,P,\rho (f))=\left( \mu -\rho (f)^{-1}\right)
^{+}$ with $\int_{0}^{1}S_{WF}(f,P,\rho (f))df=P,$ and the corresponding
rate as%
\begin{equation}
R_{WF}(P,\rho (f))=\int_{0}^{1}\left( \log _{2}\left( \mu \rho (f)\right)
\right) ^{+}df.  \label{wf}
\end{equation}%
For short, we also define $R_{WF}(P,|H(f)|^{2})=R_{WF}(P).$

In \cite{fullpaper}, the result below is proved, following \cite{ozarow}.

\textit{Lemma 1}: If $\rho (f)=|H(f)|^{2}/N$ , then we have:%
\begin{equation}
R_{WF}\left( P,\frac{|H(f)|^{2}}{N}\right) \leq \log _{2}\left( \frac{P}{N}+%
\frac{1}{1-\alpha ^{2}}\right) ,  \label{closed_upper}
\end{equation}%
where equality holds for $\frac{P}{N}\geq \frac{2\alpha }{(1-\alpha
)^{2}(1-\alpha ^{2})}.$

Having set the basic definitions above, we can now present an upper bound on
the achievable rate between the source and destination for the network in
Fig. \ref{fig1}. The bound also holds for the case where the transmitters
and receivers are informed about the codebooks used by source and
destination, and it is a straightforward consequence of cut-set arguments.

\textit{Proposition 1}: The achievable rate $R$ is upper bounded by 
\begin{eqnarray}
R_{UB} &=&\min \{C,\text{ }C^{\prime },\text{ }R_{WF}(P)\}  \label{upper} \\
&\leq &\min \left\{ C,\text{ }C^{\prime },\text{ }\log _{2}\left( P+\frac{1}{%
1-\alpha ^{2}}\right) \right\} ,  \label{upper large P}
\end{eqnarray}%
with equality in (\ref{upper large P}) for $P\geq 2\alpha /((1-\alpha
)(1-\alpha ^{2})).$

It should be noted that while the waterfilling solution (\ref{wf}) is based
on the total power constraint, due to the symmetry of the channel at hand
(see Fig. \ref{fig1}), it also satisfies the assumed per-transmitter power
constraint for any $M$ (see also \cite{somekh-review}). Moreover, the result
(\ref{upper large P}) is a consequence of Lemma 1.

\section{Finite-capacity links at the transmitter side only \label%
{sec_finite_tx}}

In this section, we consider the case in which $C$ is finite and $C^{\prime
}\rightarrow \infty $, and derive achievable rates under the assumptions
discussed above of oblivious antennas. It is noted that, due to the
infinite-capacity links at the receiver side, the assumption of oblivious
receivers has no impact on the results of this section. Two achievable rates
are derived, one that assumes knowledge at the destination of the
transmitters' codebooks and one that does not require such assumption.

\subsection{Independent messages\label{sec_nonobl_tx}}

In this section, we consider a simple scheme that assumes that the
destination is aware of the codebooks available at the transmitters. The
source splits its message (of rate $MR$) into $M$ equal-rate messages, and
delivers each to one transmitter via the finite capacity links. Each
transmitter then maps the rate-$R$ message into a codeword, using a mapping
which is known at both source and destination. The destination performs
joint decoding. It is noted that here the codebooks available at the sources
are used directly as channel codes. The following rate is achievable.

\textit{Proposition 2}: Let $C^{\prime }\rightarrow \infty .$ Then, the
following rate is achievable by transmitting independent messages (IM) from
each transmitter 
\begin{equation}
R_{IM}=\min \{C,\text{ }R_{NC}\},  \label{R_IND}
\end{equation}%
where $R_{NC}$ is the maximum rate achievable with no cooperation (NC) among
the transmitters and $C^{\prime }\rightarrow \infty $, which is given by 
\begin{align}
R_{NC}& =\int_{0}^{1}\log _{2}\left( 1+P|H(f)|^{2}\right) df=  \label{R_NC}
\\
& \hspace{-0.5in}\log _{2}\left( \frac{1+(1+\alpha ^{2})P+\sqrt{1+2(1+\alpha
^{2})P+(1-\alpha ^{2})^{2}P^{2}}}{2}\right) .  \notag
\end{align}

\textit{Remark 1}: It is easy to see that this scheme is optimal (i.e., it
achieves the upper bound (\ref{upper})) if $C\leq R_{NC}$ (and thus in
particular if $P\rightarrow \infty ).$ Instead, when $C>R_{NC}$, the rate
achievable by this scheme does not achieve the upper bound (\ref{upper}),
suffering from the performance penalty caused by independent encoding as
compared to the waterfilling solution (\ref{wf}) \cite{philosof}.

\subsection{Quantized waterfilling\label{sec_obl_tx}}

Here we consider an alternative transmission scheme in which the
transmitters' codewords are assumed to be unknown to the destination, and
thus are exploited by the source merely as quantization codebooks (of size $%
2^{nC}),$ as explained in the following. The source performs encoding for
the $M$-antenna transmitter according to the waterfilling solution (\ref{wf}%
), then it quantizes the obtained codewords using the codebooks available at
the transmitters, and send the corresponding index to the given transmitter.
Any transmitter simply transmits the codeword corresponding to the received
indices, following our assumptions. The performance of this scheme can be
proved to correspond to that of a fully cooperative MIMO system with
additional (colored) noise due to quantization, as stated in the following
proposition.

\textit{Proposition 3}: Let $C^{\prime }\rightarrow \infty .$ Then, the
following rate is achievable with quantized waterfilling (QW): 
\begin{equation}
R_{QW}=R_{WF}\left( P,\text{ }\frac{(1-2^{-C})|H(f)|^{2}}{1+P2^{-C}|H(f)|^{2}%
}\right) ,  \label{Robl_tx}
\end{equation}%
and we have:%
\begin{equation}
R_{QW}\leq \log \left( P+\frac{1}{1-\alpha ^{2}}\right) -R_{NC}(P2^{-C}),
\label{R_obltx_highSNR}
\end{equation}%
with equality in the high-SNR regime where $P\geq \frac{1}{(1-2^{-C})}\frac{%
2\alpha }{(1-\alpha )(1-\alpha ^{2})}.$

\textit{Remark 2}: The rate (\ref{R_obltx_highSNR}) reveals that for
extremely large SNR$\ (P\rightarrow \infty ),$ the rate obtained with
quantized waterfilling achieves the upper bound (\ref{upper}) $%
R_{UB}\rightarrow C$. Moreover, for large capacity $C\rightarrow \infty $,
it is easy to see from (\ref{Robl_tx}) that we have $R_{QW}\rightarrow
R_{UB}.$ This contrasts with the case of independent message transmission
studied above, where the upper bound was not achievable for large $C.$

\section{Finite-capacity links at the receive side only \label{sec_finite_rx}%
}

In this section, we focus on the scenario characterized by finite $C^{\prime
}$ and $C\rightarrow \infty .$ It is noted that, dually to the scenario
considered in Sec. \ref{sec_finite_tx}, here the assumption of oblivious
transmitters has no impact on the results. We recall that we assume
oblivious receivers in the sense specified in Sec. \ref{sec sys}. Following 
\cite{sanderovich-new}, we consider achievable rates with two quantization
strategies carried out at the receivers, in order of complexity. The first
is based on elementary compression, whereby correlation between the signals
received by different antennas is not exploited for compression, and the
second is based on distributed compression techniques. In both cases, we use
Gaussian test channels for compression.

\subsection{Elementary compression}

With elementary compression, correlation among the received signals is not
exploited in the design of the quantization functions.

\textit{Proposition 4}. Let $C\rightarrow \infty .$ Then, the following rate
is achievable with elementary compression (EC):%
\begin{equation}
R_{EC}=R_{WF}\left( \frac{P}{N_{EC}(P,C^{\prime })}\right) ,
\label{Robl_rxec}
\end{equation}%
with 
\begin{equation}
N_{EC}(P,C^{\prime })=\frac{1+(1+\alpha ^{2})P2^{-C^{\prime }}}{%
1-2^{-C^{\prime }}}.  \label{Nec}
\end{equation}%
Moreover, we have 
\begin{equation}
R_{EC}\leq \log _{2}\left( \frac{P}{N_{EC}(P,C^{\prime })}+\frac{1}{1-\alpha
^{2}}\right) ,  \label{Rec}
\end{equation}%
with equality if conditions $P\geq \frac{2\alpha }{(1+\alpha )\left(
(1+\alpha ^{2})(1-2^{-C^{\prime }})-2\alpha \right) }$ and $C^{\prime }>\log
_{2}\left( \frac{1+\alpha ^{2}}{(1-\alpha )^{2}}\right) $ are satisfied.

\textit{Remark 3}: From (\ref{Rec}), it can be seen that for extremely large
SNR$\ (P\rightarrow \infty )$ (and if the condition on $C^{\prime }$ stated
above holds)$,$ the rate achieved with elementary compression is%
\begin{equation}
R_{EC}\underset{P\rightarrow \infty }{\rightarrow }\log _{2}\left( \frac{%
2^{C^{\prime }}-1}{1+\alpha ^{2}}+\frac{1}{1-\alpha ^{2}}\right) ,
\end{equation}%
which is smaller than the upper bound (\ref{upper}) $R_{UB}\rightarrow
C^{\prime }$ for $P\rightarrow \infty .$ This shows that there is a penalty
to be paid for obliviousness at the receive side, at least if elementary
compression is employed, even when $P\rightarrow \infty $. Moreover, for
large capacity $C^{\prime }\rightarrow \infty $, we clearly have optimal
performance $R_{EC}\rightarrow R_{UB}.$

\subsection{Distributed compression}

The premise of the scheme discussed in this section is the observation that,
since decoding of all quantization codewords takes place at the destination,
the correlation of the signals observed at the receivers can be leveraged in
order to decrease the equivalent quantization noise. Following \cite%
{sanderovich-new}, the quantization scheme employed here is based on the
distributed compression approach used for the CEO\ problem.

\textit{Proposition 5}: Let $C\rightarrow \infty .$ Then, the following rate
is achievable with distributed compression (DC):%
\begin{equation}
R_{DC}=R_{WF}\left( P(1-2^{-r^{\ast }})\right) ,  \label{Roblrx}
\end{equation}%
with $r^{\ast }$ satisfying the fixed-point equation 
\begin{equation}
R_{WF}\left( P(1-2^{-r^{\ast }})\right) =C^{\prime }-r^{\ast }.
\label{fixed point}
\end{equation}%
Moreover, 
\begin{equation}
R_{DC}\leq \log _{2}\left( \frac{P+\frac{1}{1-\alpha ^{2}}}{1+P2^{-C^{\prime
}}}\right) ,  \label{rate explicit}
\end{equation}%
with equality if conditions $P\geq \frac{2\alpha }{(1-\alpha )\left(
(1-\alpha ^{2})-2^{-C^{\prime }}\right) }$ and $C^{\prime }>2\log _{2}\left( 
\frac{1}{1-\alpha }\right) $ are satisfied.

\textit{Remark 4}: Equation (\ref{fixed point}) is easily solved numerically
since $R_{WF}\left( P(1-2^{-r^{\ast }})\right) $ is a monotonic function of $%
r^{\ast }.$ Moreover, the expression (\ref{rate explicit}) shows that for
extremely large SNR$\ (P\rightarrow \infty )$, the rate with oblivious
transmitters (if the condition on $C^{\prime }$ given above is satisfied,
which requires sufficiently small $\alpha $ or large $C^{\prime }$) achieves
the upper bound (\ref{upper}), i.e., $R_{DC}\rightarrow C^{\prime }.$ This
contrasts with the result discussed in\ Remark 3 for elementary compression,
which was shown to be unable to achieve the upper bound. Finally, it can be
seen that for large capacity $C^{\prime }\rightarrow \infty $, we have $%
R_{DC}\rightarrow R_{UB}.$

\section{Finite-capacity links at the transmit and receive sides\label%
{sec_finite_both}}

In the two previous sections, we have considered the two limiting cases $%
C^{\prime }\rightarrow \infty $ (Sec. \ref{sec_finite_tx}) and $C\rightarrow
\infty $ (Sec. \ref{sec_finite_rx}), and constructed basic transmission and
reception strategies based on oblivious antennas, namely, transmission of
independent messages (IM) versus quantized waterfilling (QW) at the transmit
side, and elementary (EC)\ versus distributed compression (DC) at the
receive side. These techniques can be combined giving rise to four
transmission/ reception strategies (IM-EC, IM-DC, QW-EC and QW-DC), as
discussed below.

\subsection{Independent messages and elementary compression}

It is recalled that, when using transmission of independent messages, it is
assumed that the destination is aware of the codebooks available at the
transmitters.

\textit{Proposition 6}: The following rate is achievable by transmitting
independent messages (IM) and using elementary compression (EC) at the
receive side:%
\begin{equation}
R_{IM-EC}=\min \left\{ C,\text{ }R^{\prime }\right\} ,  \label{IMEC}
\end{equation}%
with $R^{\prime }=\log _{2}\left( \frac{N_{EC}+(1+\alpha ^{2})P+\sqrt{%
(N_{EC}+(1+\alpha ^{2})P)^{2}-4\alpha ^{2}P^{2}}}{2N_{EC}}\right) $ and $%
N_{EC}(P,C^{\prime })$ as in (\ref{Nec}) (we have dropped the dependence on $%
P,$ $C^{\prime }$ for the sake of legibility).

\textit{Remark 5}: The result in Proposition 2 can be found as a special
case of Proposition 6 for $C^{\prime }\rightarrow \infty .$ Moreover, this
scheme is optimal whenever the second term in (\ref{IMEC}) is larger than $%
C. $ For $P\rightarrow \infty ,\ $as shown in \cite{fullpaper}, the scheme
is not optimal and when $C,$ $C^{\prime }\rightarrow \infty $, the we have $%
R_{IM-EC}\rightarrow R_{NC}\leq R_{UB},$ thus suffering from the performance
loss due to transmission of independent messages (see\ Remark 2).

\subsection{Independent messages and distributed compression}

\textit{Proposition 7}: The following rate is achievable by transmitting
independent messages (IM) and using distributed compression (DC) at the
receive side:%
\begin{equation}
R_{IM-DC}=\min \{C,\text{ }R^{\prime }\}  \label{R_IMDC}
\end{equation}%
with $R^{\prime }=\log _{2}\left( \frac{1+AP+2\alpha ^{2}2^{-C^{\prime
}}P^{2}+\sqrt{1+2AP+\left( B^{2}+4\alpha ^{2}2^{-C^{\prime }}\right) P^{2}}}{%
2(1+2^{-C^{\prime }}P)(1+\alpha ^{2}2^{-C^{\prime }}P)}\right) $ ($%
A=(1+\alpha ^{2})$ and $B=(1-\alpha ^{2})).$

\textit{Remark 6}: Proposition 2 follows from Proposition 7 when $C^{\prime
}\rightarrow \infty .$ Moreover, as for the previous scheme, optimality is
guaranteed if the second term in (\ref{R_IMDC}) is larger than the capacity $%
C$. However, optimality is also attained with $P\rightarrow \infty $ (see
also Remark 4), while for $C$ and $C^{\prime }\rightarrow \infty ,$ we have $%
R_{IM-DC}\rightarrow R_{NC}\leq R_{UB}.$

\subsection{Quantized waterfilling and elementary compression}

We recall that, unlike the previous two subsections, the scheme considered
here, based on quantized waterfilling, does not require the destination to
be aware of the codebooks available at the transmitters.

\textit{Proposition 8}: The following rate is achievable by using quantized
waterfilling (QW) at the transmit side and elementary compression (EC) at
the receive side:%
\begin{equation}
R_{QW-EC}=R_{WF}\left( P,\text{ }\frac{(1-2^{-C})|H(f)|^{2}}{%
N_{EC}(P,C^{\prime })+P2^{-C}|H(f)|^{2}}\right) ,  \label{R_obl}
\end{equation}%
with $N_{EC}(P,C^{\prime })$ as in (\ref{Nec}). An upper bound and large-$P$
closed-form expression for (\ref{R_obl}) can be found in \cite{fullpaper}.

\subsection{Quantized waterfilling and distributed compression}

\textit{Proposition 9}: The following rate is achievable by using quantized
waterfilling (QW) at the transmit side and distributed compression (DC) at
the receive side:%
\begin{equation}
R_{QW-DC}=R_{WF}\left( P,\text{ }\frac{(1-2^{-r^{\ast }})(1-2^{-C})|H(f)|^{2}%
}{1+P2^{-C}|H(f)|^{2}}\right)  \label{R_obldc}
\end{equation}%
with $r^{\ast }$ satisfying the fixed-point equation 
\begin{equation}
R_{WF}\left( P,\text{ }\frac{(1-2^{-r^{\ast }})(1-2^{-C})|H(f)|^{2}}{%
1+P2^{-C}|H(f)|^{2}}\right) =C^{\prime }-r^{\ast }.  \label{fpt}
\end{equation}%
An upper bound and a large-$P$ closed-form expression can be found in \cite%
{fullpaper}.

\textit{Remark 7: }While Proposition 8 subsumes Propositions 3 for $%
C^{\prime }\rightarrow \infty $ and 4 for $C\rightarrow \infty ,$
Proposition 9 entails Proposition 3 for $C^{\prime }\rightarrow \infty $ and
Proposition 5 for $C\rightarrow \infty .$ Moreover, equation (\ref{fpt}) is
easily solved numerically since the left-hand side is a monotonic function
of $r^{\ast }.$ Finally, reference \cite{fullpaper} shows that, for both
QW-DC and QW-DC, for $P\rightarrow \infty $ the upper bound is not attained,
while when $C$ and $C^{\prime }\rightarrow \infty $ the opposite is true.

\section{Numerical results\label{sec_numerical}}

\begin{figure}
\centering
 \includegraphics[scale=0.45]{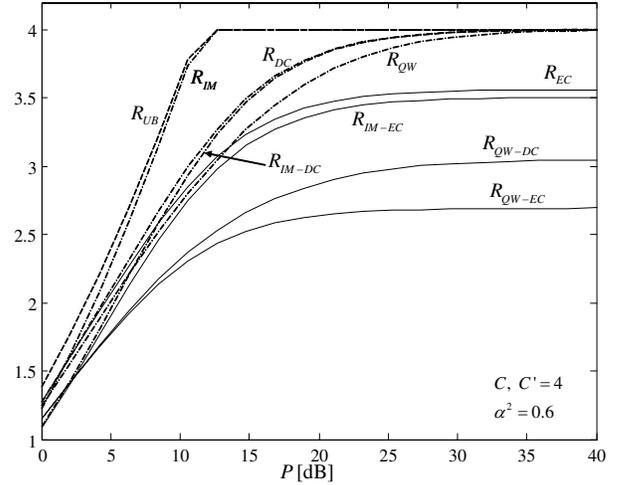}
 \caption{Achievable rates versus SNR\ $P. $ Rates $R_{IM}$ and $R_{QW}$ are for $C^{\prime }\rightarrow \infty $ and $C=4,$ rates $R_{EC}$ and $R_{DC}$ to $C\rightarrow \infty $ and $C^{\prime }=4,$ while the remaining curves correspond to $C=C^{\prime }=4$ ($\protect\alpha ^{2}=0.6$).}
 \label{fig2}
\end{figure}

Fig. \ref{fig2} shows the achievable rates of interest versus the SNR\ $P$
for $\alpha ^{2}=0.6$. Starting with the case $C\rightarrow \infty $ of Sec. %
\ref{sec_finite_tx} and $C^{\prime }=4$, it is noted that in this scenario,
exploiting knowledge of the transmitters' codebooks at the destination via
independent encoding ($R_{IM}$) enables the upper bound $R_{UB}$ to be
closely approached and attained for sufficiently large SNR, here $P\gtrsim
10dB$ (see Remark 1). The use of quantized waterfilling, instead, allows the
upper bound to be achieved only for extreme SNR; here $P\gtrsim 40dB$ (see
Remark 2). For the case $C^{\prime }\rightarrow \infty $ of Sec. \ref%
{sec_finite_rx} and $C=4$, it is concluded that, while distributed
compression is able to achieve the upper bound $R_{UB}$ for $P\rightarrow
\infty $, the same is not true for elementary compression (see Remarks 3 and
4). Similar conclusions carry over to the case of finite $C$ and $C^{\prime
} $ ($C=C^{\prime }=4)$: for instance, with large power $P$, the upper bound
can be reached only if independent messages are transmitted with distributed
compression (see\ Remark 6). Moreover, distributed compression significantly
outperforms elementary compression, especially for high power $P.$


\section{Concluding remarks\label{sec concl}}

A distributed MIMO scenario with transmit and receive antennas that are
oblivious to the codebook of source and destination has been considered.
Achievable rates have been derived based on several proposed techniques that
exploit both channel and source coding principles. Referring the reader to 
\cite{fullpaper} for a full discussion, here we point out that the analysis
has shown that the considered design with oblivious antennas does not entail
any loss of optimality in specific asymptotic and non-asymptotic regimes of
SNR and link capacities. These results are in accord with the conclusions of
recent work reviewed in \cite{shamai-review} for uplink and downlink
channels with finite-capacity backhaul.

\section*{Acknowledgment}

{}This research was supported in part by a Marie Curie Outgoing
International Fellowship and the NEWCOM++ network of excellence both within
the 7th European Community Framework Programme, by the REMON Consortium and
by the U.S. National Science Foundation under Grants CNS-06-25637,
CNS-06-26611, and ANI-03-38807.


\bigskip

\bigskip

\bigskip

\bigskip

\bigskip

\bigskip

\bigskip

\bigskip

\bigskip

\bigskip

\bigskip

\bigskip

\bigskip

\bigskip

\bigskip

\bigskip

\end{document}